\documentclass[sigconf]{acmart}

\AtBeginDocument{%
  }

\usepackage[utf8]{inputenc}
\usepackage{tabularx}
\usepackage{booktabs}
\usepackage{soul}
\usepackage{xcolor}
\usepackage{xspace}
\usepackage{graphicx}
\usepackage[caption=false]{subfig}  
\usepackage[export]{adjustbox}
\usepackage{amsmath}
\usepackage{textcomp}
\usepackage{listings}
\usepackage{cleveref}
\usepackage{algorithm}
\usepackage[noend]{algpseudocode}
\usepackage{balance}
\usepackage{enumitem}
\usepackage{multirow}
\usepackage{colortbl}
\usepackage{balance}
\usepackage{threeparttable}
\usepackage{multicol}
\usepackage{tcolorbox}

\definecolor{codegreen}{rgb}{0,0.6,0}
\definecolor{codegray}{rgb}{0.5,0.5,0.5}
\definecolor{codepurple}{rgb}{0.58,0,0.82}
\definecolor{backcolour}{rgb}{0.95,0.95,0.92}
\definecolor{gray}{gray}{0.9}
\definecolor{APA_stats}{RGB}{100, 100, 120}

\newcounter{observation}

\setcopyright{acmcopyright}
\copyrightyear{2023}
\acmYear{2023}
\acmDOI{XXXXXXX.XXXXXXX}

\acmConference[ICSE 2024]{46th International Conference on Software Engineering}{April 2024}{Lisbon, Portugal}
\acmPrice{15.00}
\acmISBN{978-1-4503-XXXX-X/18/06}

\begin{document}

\title{Embedding-based search in JetBrains IDEs}


\author{Evgeny Abramov}
\email{evgeny.abramov@jetbrains.com}
\affiliation{%
  \institution{JetBrains}
  \city{Limassol}
  \country{Cyprus}
}
\author{Nikolai Palchikov}
\email{nikolai.palchikov@jetbrains.com}
\affiliation{%
  \institution{JetBrains}
  \city{Munich}
  \country{Germany}
}

\renewcommand{\shortauthors}{Abramov et al.}

\begin{abstract}
Most modern Integrated Development Environments (IDEs) and code editors have a feature to search across available functionality and items in an open project. In JetBrains IDEs, this feature is called Search Everywhere: it allows users to search for files, actions, classes, symbols, settings, and anything from VCS history from a single entry point. However, it works with the candidates obtained by algorithms that don’t account for semantics, e.g., synonyms, complex word permutations, part of the speech modifications, and typos. In this work, we describe the machine learning approach we implemented to improve the discoverability of search items. We also share the obstacles encountered during this process and how we overcame them.

\end{abstract}





\maketitle

\section{Introduction}
Many modern code editors feature a search function that allows users to navigate through available features and various components of a project. One can conduct this search in several ways, including seeking exact matches.

Nevertheless, challenges arise when users struggle to recall the precise contents of the element they are searching for.
This issue prompts exploring an improved search function that is robust enough to consider synonyms, manage typos, and operate speedily on local computers.

JetBrains integrated development environments (IDEs) utilize a heuristic algorithm to make the search more flexible. However, it has some limitations as it is based on regular expressions and does not consider aspects like word permutation in a query or synonym recognition. Modern search implementations commonly use semantic vector representations of searchable entities. This observation leads us to investigate how we can improve our search with embedding-based methods.

\begin{figure}[t]
  \centering
  \captionsetup{justification=centering}
  \includegraphics[width=\columnwidth]{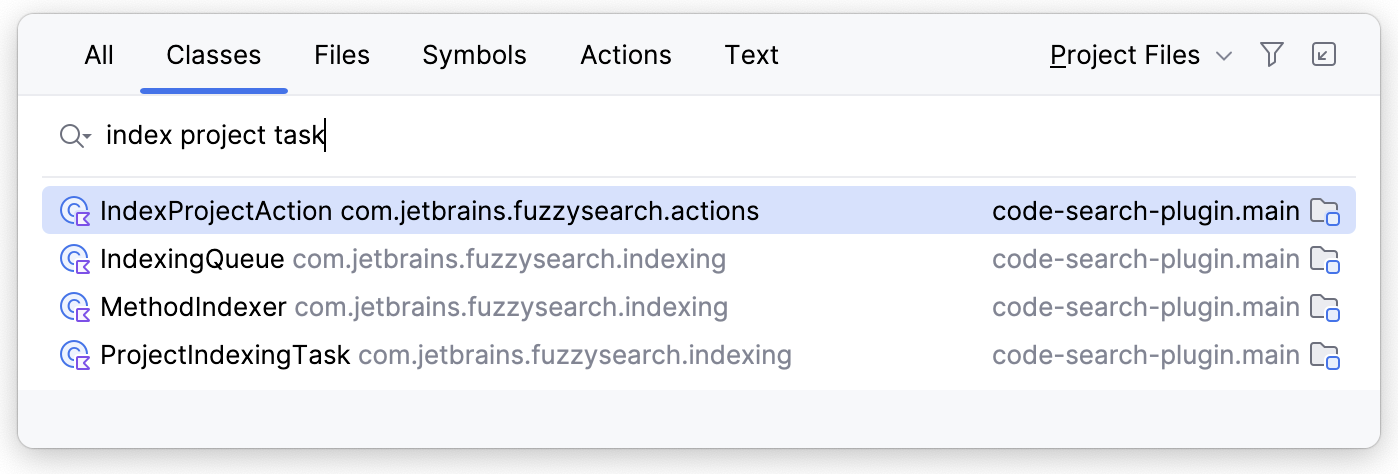}
  \caption{Search Everywhere dialog with\\enabled embedding-based search}
  \label{fig:ide_screenshot}
  \vspace{-10pt}
\end{figure}

Let us first consider the context of the problem and the requirements. JetBrains IDEs are desktop Java Virtual Machine (JVM) applications developed primarily with Java and Kotlin. Our desire for applications to operate effectively on any hardware imposes significant limitations on our employment of available algorithms. We initially experimented with a server-based approach to eliminate this restriction, allowing us more freedom in choosing hardware and software stacks. We encountered a few critical drawbacks with it, so we later moved to an entirely local approach that resolved most of them. First, we could not deliver the search to all users because of the queries and code privacy we generally provide. Second, for users who agree to send the data, the required resources to store and operate the appropriate versions of project indices would be tremendous. And third, the improved search would require a stable internet connection and additional requests round trip time. With our current method, we precalculate vector representations for all indexable items on users' computers and leverage these vectors to determine the relevance between a query and an item during the search. Our approach minimizes the required neural network inference time during the search, making its latency comparable to the standard search. We now provide this experimental functionality to the users of the Early Access Program of JetBrains IDEs. One can see an example of embedding-based retrieval in Search Everywhere on the screenshot, depicted in Figure \ref{fig:ide_screenshot}.

\section{Related work}

We used the results from the Sentence-BERT paper \cite{reimers-2019-sentence-bert}, one of the first to propose using a bi-encoder Siamese architecture with the same model applied separately to queries and items. During the training procedure for the metric learning task, we considered the loss functions introduced in ``Supervised Contrastive Learning'' \cite{khosla2020supervised} and ``Learning a distance metric from relative comparisons'' \cite{schultz2003learning}. We train our models on datasets introduced in the ``CodeSearchNet Challenge'' \cite{husain2019codesearchnet} and ``CoSQA'' \cite{huang2021cosqa}. To integrate the models with IDEs, we converted them to ONNX \cite{bai2019}, an open-source format for AI models that defines the computational graph structure, intermediate operators, and data types. We then used the KInference \cite{kinference} library, which allows us to run machine learning models in Kotlin with minimal additional dependencies. We were inspired to use smaller models by the ``Well-Read Students Learn Better: On the Importance of Pre-training Compact Models'' \cite{turc2019well}.
\section{Our approach}
We apply the language model to some text associated with an indexable entity to calculate a vector representation. For actions, the text representation is an action name that users observe in the search results. By splitting the names of corresponding items into words with regular expressions, we reduce the search tasks for files, classes, and methods to natural language search.
 It is possible since users usually name items according to the snake or camel case, with relatively few exceptions. We also considered using more specific descriptions of actions and the complete source code of methods as textual representations of items. Despite improved search quality, this notably slowed the indexing duration, so we dismissed these options for now.
\begin{figure}[h]
  \centering
  \includegraphics[width=\columnwidth]{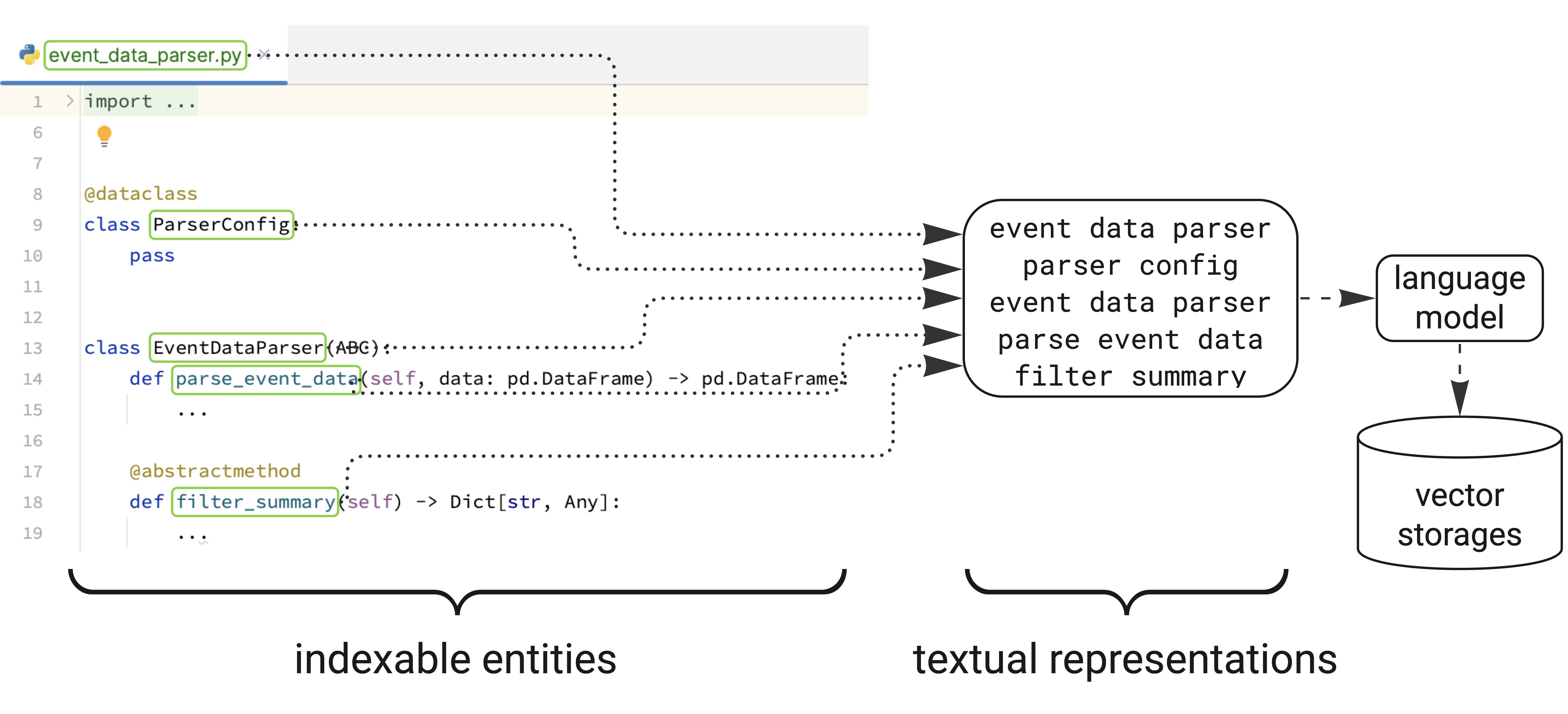}
  \caption{Indexing process}
  \label{fig:indexing}
\end{figure}

Embeddings for indexable items are stored on disk consecutively. Fixed embedding vector size allows us to do random access over the embedding storage and synchronize every incremental change operation with the disk without rewriting the whole file. The overall scheme of how indexing works is depicted in Figure \ref{fig:indexing}.

\begin{figure}[h]
  \centering
  \includegraphics[width=\columnwidth]{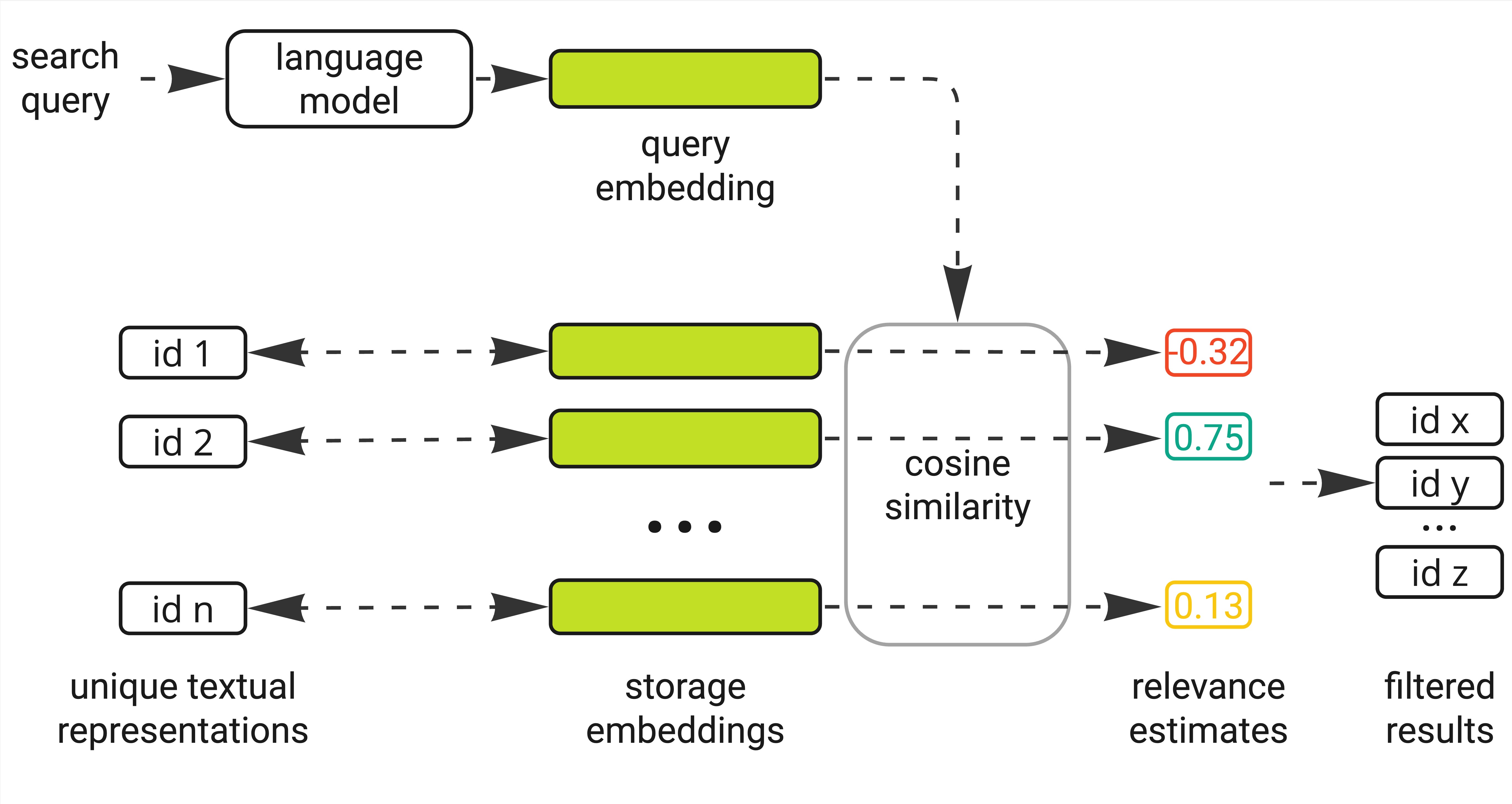}
  \caption{Search process}
  \label{fig:pipeline}
\end{figure}

As the distance metric during the search, we use cosine similarity between precalculated vectors in storage and vector representation of a query from the neural language model. Despite time-effective approximate nearest neighbors search algorithms, we use brute force iteration over embeddings for simplicity. With this approach, we can stream the results and display found items before completing the search.
We discovered that dynamically changing the similarity threshold is very beneficial to decide whether an item is relevant depending on the number of found items.
The overall approach is depicted in Figure \ref{fig:pipeline}.

We use a single model to search over all types of items to reduce the memory footprint during the runtime and the size of the model artifacts on disk. We convert the model weights to 16-bit floating point format for the same purpose.

We perform semantic search concurrently to not slow down the standard search. Existing functionality of all IDE subsystems is available regardless of vector indices, so users do not have to wait longer to start working on a project.

Search Everywhere uses the CatBoost gradient boosting for ranking. To integrate with this functionality, we introduced a new ranking feature, the cosine similarity between the query and item embeddings. Until we train the ranking models to handle cosine similarity properly, we place the suggestions from the embedding search at the end of the search results list.

\subsection{Model}
We considered several architectures of language models. As a starting point, we integrated the MiniLM \cite{wang2020minilm} architecture with six attention layers, an attention-based network trained from a larger teacher model. Despite the reasonable search time, embedding calculations for indexable items could take a few hours for large projects, so we considered smaller models and gave up using attention layers. We discovered that even tiny models consisting of an embedding layer and two linear layers perform well when fine-tuned on datasets specific for actions and code search tasks. We reduced the indexing time to a few minutes with a lower model size, even for large projects with hundreds of thousands of files. Table \ref{tab:latency} demonstrates the average one item indexing duration of three model architectures applied to the IntelliJ Community repository \cite{intellij}. It also reveals the difference in memory and disk requirements for model inference and storage. As another optimization, we significantly reduced the memory required to store a project index after changing the embedding size from 384 to 128.

\begin{table}[h!]
    \centering
    \captionsetup{justification=centering}
    \resizebox{\columnwidth}{!}{%
    \begin{tabular}{m{2.5cm}ccc}
        & \textbf{Our model} & \textbf{SentenceBERT} & \textbf{E5-small-v2} \\
        \\[-1em]
        \hline
        \\[-1em]
        Average item\newline indexing duration & 0.5 ms & 14.8 ms & 29.2 ms \\
        \\[-1em]
        \hline
        \\[-1em]
        Computation graph\newline memory footprint & 22 MB & 104 MB & 157 MB \\
        \\[-1em]
        \hline
        \\[-1em]
        Model weights\newline disk footprint & 9 MB & 47 MB & 69 MB \\
    \end{tabular}%
    }
    \smallskip
    \caption{Indexing performance and storage requirements comparison}
    \label{tab:latency}
    \vspace{-23pt}
\end{table}

\begin{table*}[t]
    \centering
    \resizebox{0.7\textwidth}{!}{%
        \begin{tabular}{l|cc|cc|cc}
            & \multicolumn{2}{c|}{\textbf{Our model}} & \multicolumn{2}{c|}{\textbf{Pre-trained SentenceBERT}} & \multicolumn{2}{c}{\textbf{Pre-trained E5-small-v2}} \\
            \textbf{Dataset} & \textbf{NDCG@10} & \textbf{MRR@10} & \textbf{NDCG@10} & \textbf{MRR@10} & \textbf{NDCG@10} & \textbf{MRR@10} \\
            & \multicolumn{2}{c|}{} & \multicolumn{2}{c|}{} & \multicolumn{2}{c}{}\\[-1em]
            \hline
            \\[-1em]
            \textit{CodeSearchNet} & 0.5461 & 0.5975 & 0.5406 & 0.5297 & 0.5500 & 0.5893 \\
            \textit{CoSQA} & 0.7213 & 0.6662 & 0.5278 & 0.4766 & 0.4684 & 0.4150 \\
            \textit{JetBrains Actions} & 0.9062 & 0.8925 & 0.9038 & 0.8800 & 0.8868 & 0.8681 \\
        \end{tabular}%
    }
    \smallskip
    \caption{Metrics comparison of our model vs. pre-trained sentence embedding models\\(w/o downstream task fine-tuning)}
    \label{tab:performance}
    \vspace{-10pt}
\end{table*}

\subsection{Training}
We train our models on the concatenation of two datasets. The first is our labeled dataset with the pairs of queries and names of relevant IDE actions. The second contains pairs of queries and relevant methods from the CoSQA dataset. We trained our current production model with the contrastive loss version with similarity metric where positive and negative margins were equal to 1 and 0, respectively.

\subsection{Evaluation}
For the offline evaluation, we used normalized discounted cumulative gain (NDCG) \cite{jarvelin2002cumulated} and mean reciprocal rank (MRR) \cite{voorhees1999trec} to measure the ranking quality. Table \ref{tab:performance} demonstrates that our model performs comparably to pre-trained SentenceBERT and E5-small-v2 \cite{wang2022text} models, which we did not fine-tune on the datasets.

\begin{figure}[t]
  \centering
  \captionsetup{justification=centering}
  \includegraphics[width=\columnwidth]{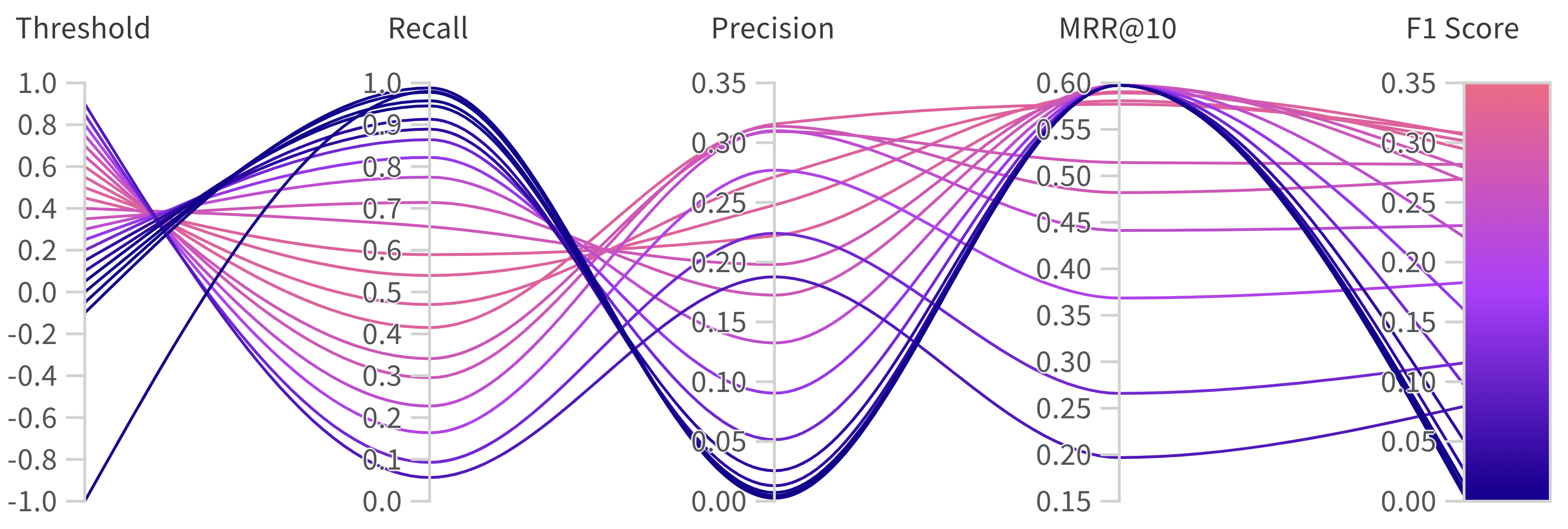}
  \caption{Parallel coordinates plot to compare multiple metrics for several similarity thresholds}
  \label{fig:sweeps}
\end{figure}

\begin{table}[t]
    \centering
    \captionsetup{justification=centering}
    \resizebox{\columnwidth}{!}{%
    \begin{tabular}{l|cc|cc}
        & \multicolumn{4}{c}{\textbf{Textual representation}} \\
        \\[-1em]
        & \multicolumn{2}{c|}{\textbf{Function name}} & \multicolumn{2}{c}{\textbf{Function body}} \\
        \textbf{Dataset} & \textbf{NDCG@10} & \textbf{MRR@10} & \textbf{NDCG@10} & \textbf{MRR@10} \\
        & \multicolumn{2}{c|}{} &\multicolumn{2}{c}{}\\[-1em]
        \hline
        \\[-1em]
        \textit{CodeSearchNet} & 0.5406 & 0.5297 & 0.6191 & 0.5927 \\
        \textit{CoSQA} & 0.5278 & 0.4766 & 0.8354 & 0.7881 \\
    \end{tabular}%
    }
    \smallskip
    \caption{Importance of textual representation\\for functions search quality}
    \label{tab:functions}
    \vspace{-23pt}
\end{table}

With the above-mentioned ranking metrics and relevance classification metrics, we selected the similarity threshold to determine the relevance of the result. We used a parallel coordinates plot, like in the figure \ref{fig:sweeps}, to manually compare multiple metrics for several similarity thresholds. Each curve represents one threshold value and demonstrates the corresponding filtering and ranking capabilities.

Table \ref{tab:functions} illustrates the search quality benefit of indexing whole function bodies instead of only names (the metrics are calculated for the SentenceBERT model).

\section{Future Work \& Open Questions}

Several questions influence the production-readiness of the software feature outlined herein. Each of the following questions not only describes a possible direction of our future work but also extends an invitation to the research community for active contributions.

\begin{enumerate}
    \item There is a trade-off between the inference speed and quality of embeddings. The question is \textit{How to balance it properly, considering all details (for example, the user's machine)}. Possible work might be refining different techniques for model optimization (quantization/distillation/pruning) or for inference algorithm (for example, HNSW. \cite{malkov2018efficient}).
    \item \textit{What is the optimal way of injecting searchable item context into an embedding}? There are different approaches (for example, \cite{orvieto2023resurrecting}, \cite{beltagy2020longformer}, and many more) for training models on long sequences and extended contexts. One possible methodology could involve fine-tuning these models using metric learning, although real-time updating of the vector index might be challenging in numerous cases. Naturally, online evaluations will have the final say. There is a possibility that including a more extended context will not necessarily improve task performance, as might be shown by online metrics.
    
    \item We can choose from different context sizes. The question then is: \textit{What minimum context size still ensures a high-quality embedding?}
\end{enumerate}

We eagerly invite the community to help guide our journey towards understanding these issues more thoroughly.
\section{Conclusion}

In this paper, we are discussing our ongoing work related to the adoption of an embedding-based search feature within the Integrated Development Environment. At this stage, we have created a robust baseline model which is efficient in terms of indexing speed. This makes it sufficient for practical applications in real-time scenarios.

Nevertheless, there are plenty of opportunities to refine this model. An example of this could be taking larger contexts into account when calculating embeddings for a specific item. Simultaneously, while we are evaluating our prototype online, we have pinpointed the key problems that we aim to resolve in the near future.

In essence, this paper provides a snapshot of our current progress and maps out the key objectives for the ongoing development of this crucial feature.


\bibliographystyle{ACM-Reference-Format}
\bibliography{main_refs}


\end{document}